\documentclass[12pt]{iopart}
\usepackage{graphicx}
\usepackage{amssymb}

\begin{document}
\rapid{Large field generation with Hot Isostatically Pressed Powder-in-Tube MgB$_2$ coil at 25 K}

\author{A. Serquis\dag \footnote[2]{on leave from CONICET - Centro At\'omico Bariloche, 8400, S.C. de Bariloche, Argentina.}, L. Civale \dag, J.~Y. Coulter\dag, D.~L. Hammon\dag, X.~Z. Liao\dag, Y.~T. Zhu\dag, D.~E. Peterson\dag, F.~M. Mueller\dag, V.~F. Nesterenko\S, and S.~S. Indrakanti\S}

\address{\dag Superconductivity Technology Center, MS K763, Los Alamos National Laboratory, Los Alamos, NM 87545, USA}
\address{\S Department of Mechanical and Aerospace Engineering, University of California, San Diego, La Jolla, CA 92093, USA}

\ead{aserquis@cab.cnea.gov.ar}

\begin{abstract} 
We present the fabrication and test results of Hot-Isostatic-Pressed (HIPed) Powder-in-Tube (PIT) 
MgB$_2$ coils. The coils properties were measured by transport and magnetization at 
different applied fields ($H$) and temperatures ($T$).
The engineering critical current ($J_e$) value is the largest reported in PIT MgB$_2$ wires
or tapes.  At 25 K our champion 6-layer coil was able to generate a field of 1 T at 
self-field ($I_c >$ 220 A, $J_e \sim 2.8 \times 10^4$ A/cm$^2$). 
At 4 K this coil generated 1.6 T under an applied field of 1.25 T ($I_c \sim350$ A, $J_e \sim 
4.5 \times 10^4$ A/cm$^2$). These magnetic fields are high enough for a superconducting transformer 
or magnet applications such as MRI. A SiC doped MgB$_2$ single layer coil shows a promising 
improvement at high fields and exhibits $J_c > 10^4$ A/cm$^2$ at 7 T. 
 
\end{abstract}

\pacs{74.70.Ad, 74.60.Jg, 74.62.Bf}


\maketitle

Soon after the discovery of superconductivity in MgB$_2$ \cite{nagamatsu2001}, it became 
clear that it has strong potential for commercial applications due to a unique combination of 
characteristics, such as the high transition temperature $T_c \sim 39~$K, the chemical 
simplicity, and the low cost of raw materials. In addition, the absence of weak-link 
behavior at grain boundaries in polycrystalline samples \cite{bugovslavsky2001} permits 
the use of simple powder in tube (PIT) methods to fabricate wires \cite{jin2001}. Indeed 
PIT wires with reasonably good properties were fabricated early 
on \cite{SuoAPL,Fujii,Soltanian,Flukiger}, making MgB$_2$ a candidate to replace 
NbTi or Nb$_3$Sn in magnets. 
The main applications require high $J_c$ values at high fields and intermediate 
temperatures, which can be reached by both liquid He as well as commercial cryocoolers. 
However, the high critical current densities $J_c$ ($\sim10^6$ A/cm$^2$) usually reported
 at zero field and $\sim 4$ K  for MgB$_2$ wires or tapes \cite{Flukiger} rapidly decrease 
with increasing temperature or 
magnetic field. To increase $J_c$, it is necessary to introduce more pinning centers 
and also overcome the poor connectivity between grains \cite{serquis2003a}. Besides, 
the figure of merit of a superconductor not only depends on $J_c$, but also on the 
engineering critical current density $J_e$ = $fJ_c$, where $f$ is the filling factor 
that reflects the relative ratio of the superconductor to the total cross section of 
the wire. In this work we present the results of the optimization of several processing 
parameters to fulfill these requirements not only in short wires but also in the 
fabrication of coils.\\

First, an adequate and inexpensive sheath material must be selected. We packed MgB$_2$ 
powder into stainless steel tubes (inner and outer diameters 4.6 and 6.4 mm) and cold-drew 
them into round wires with external diameter in the range of  0.8-1.4 mm, with one 
intermediate annealing. This results in MgB$_2$ cores of very uniform, circular cross 
section, with diameters of 0.5-0.9 mm corresponding to $f\sim 45\%$ and no reaction 
between the sheath and the superconductor \cite{serquis2003a,serquis2003b}.\\

A second requirement is to obtain a good inter-grain connectivity by precluding excessive 
porosity and microcracks, and the formation of large non-superconducting precipitates, 
such as MgO, at grain boundaries. We have shown \cite{serquis2003a,serquis2003b} that, by 
adding 5 at.$\%$ Mg to the initial powder, heat-treating long lengths of wire with sealed 
ends to avoid Mg loss, and by choosing appropriate annealing conditions, the microcracks 
produced by the drawing (the most severe current-limiting factor in the as-drawn wires) can 
be healed due to a recrystallization promoted by the excess Mg.\\

Finally, we have found \cite{serquis2003c,liao2003a}  that to improve $J_c$ at high magnetic 
fields ($H$) and temperatures ($T$), our hot isostatic pressing (HIP) of the 
PIT wires produces significantly better results than ambient-pressure annealings. 
In addition to eliminating most of the MgO precipitates, the microcracks, and the porosity, 
HIPing introduces a high density of crystalline structural defects, including small angle twisting, 
tilting, and bending boundaries. This results in the formation of subgrains within the 
MgB$_2$ crystallites with a high dislocation density at subgrain boundaries \cite{liao2003a}.
 These additional pinning centers produce a $J_c$ enhancement (with respect to the 
ambient-pressure annealings) that is marginal at $T=4~$K and self-field, but becomes very 
significant as $T$ and $H$ increase, e.g., a factor of $\sim 4$ at $T=26~$K and 
$\mu_0H=2~$T \cite{serquis2003c}. These short wires achieved the highest reported 
$H_{irr}$ for PIT MgB$_2$ wires, up to $\sim 17~$T at $T=4~$K. On the other hand, nano-sized
 SiC doped MgB$_2$ wires, prepared by Dou \textit{et al.} \cite{dou2002}, also show a 
significant enhancement of critical current density in high magnetic fields over a wide 
temperature range. Hence, it is interesting to investigate the combined effect of HIPing 
on SiC doped MgB$_2$ wires.\\

It is necessary to demonstrate that the high-quality properties mentioned above can be 
achieved in longer wires. A few groups have reported the fabrication of long wires or 
tapes for the construction of MgB$_2$ coils or magnets \cite{coils}, but the $J_e$ results 
are below the values corresponding to short wires.
Therefore, we decided to prepare HIPed MgB$_2$ coils capable to produce magnetic fields 
useful for applications such as MRI, at temperatures compatible with liquid-helium-free 
operation. Below we report the characteristics and performance of our best coils.\\

We wound 25 m of our 1 mm diameter, as-drawn PIT wire, around a 3 cm-diameter stainless 
steel barrel, into a 3.1 cm-long, 4.5 cm-external diameter, 6-layer coil, with insulating 
fiber-glass fabric intercalated in between layers. The coil was fixed on the outside 
surface of the barrel to avoid large strain deformation of the wire on the stage of 
cooling and pressure release due to the difference in coefficients of thermal expansion 
and elastic moduli of the wire core and the sheath.  These strains may result in the 
fracture of the brittle high density magnesium diboride core. The set was HIPed at 900$^{\circ}$C 
under a pressure of 200 MPa for 30 min, depressurized, and cooled at 5$^{\circ}$C/min 
to room $T$. For measurements, we removed the coil from the barrel, made $\sim 15~$cm 
long current contacts by soldering a Cu tape, added two voltage contacts (placed about 
24 m apart on the wire) to measure the I-V curves, attached a Hall probe at the coil 
center, and inserted the coil coaxially in the 5 cm bore of a 9 T superconducting magnet.\\

We tested the coil both immersed in liquid He (4 K) and in liquid Ne (24.6 K $<T<$ 26.5 K). 
Fig. 1 shows (in solid symbols) the maximum current that we were able to put through the 
coil as a function of applied field ($H^{a}$). At 4 K a break in the $I_c(H^{a})$ curve 
is visible at $\sim 7~$T, which coincides with a clear change in the I-V curves. Above 7 T 
the I-V curves were smooth and had a well behaved n-value, indicating that we were correctly
 measuring $I_c$, while at lower $H^{a}$ the voltage suddenly jumped from almost zero to a 
large value, implying a quench of the coil due to the heat propagating from the current 
contacts. At 25 K (pumped liquid Ne) we were always in that situation (regardless of this, 
for simplicity we call the maximum current $I_c$ in all cases). The solid and dotted lines 
are the $I_c$ obtained from magnetization measurements using the Bean model on a 0.5 cm 
long piece of the wire. 
The excellent coincidence of the magnetization and transport $I_c$ at 4 K above 7 T 
indicates that the 25 m long wire is very homogeneous. Indeed, the performance is as good 
as that of our short HIPed wires previously reported \cite{serquis2003c}.\\

For each $(H^{a},T)$ in Fig. 1, we used the Hall probe to measure the total field at 
the coil center when a current $I_c$ was applied (open symbols). The field difference 
between the open and solid symbols at a given $I_c$ and $T$ (connected by arrows in the 
figure) is the field generated by the coil ($H^{gen}$). We found excellent proportionality 
$H^{gen}=(4.5~$mT/A$)*I_c$ (dashed line in fig. 1). At 4 K we were limited by the maximum 
current of the power supply, 350 A, which generated a field $H^{gen}=1.6~$T at 
$H^{a}=1.25~$T, resulting in a total field at the coil center of 2.85 T and corresponds 
to a $J_e \sim 4.5 \times 10^4 A/cm^2$. The intersection of the solid and dashed lines 
indicate that at 4 K and $H^{a}=0$ this coil could generate a field of $\sim 2.5~$T, 
at a current of $\sim 550~$A.\\

We also measured $I_c(T)$ and $H^{gen}(T)$ (with the Hall probe) at $H^{a}=0$, 
by pumping the liquid Ne, as shown in Fig. 2. At $T=26.5~$K (ambient pressure at Los Alamos),
$H^{gen}=0.85~$T, and at $T=24.6~$K (slightly above the Ne triple point) 
$H^{gen}(24.6~$K$)=1$ T. This correspond to a $J_e \sim 2.8 \times 10^4 A/cm^2$.\\ 

Another single layer coil was made following the same procedure with a 4 meter length wire 
that contains MgB$_2 + 5$ \% SiC, to explore the effect of microprecipitates in combination 
with the HIPing process.  Fig. 3 displays  (in solid circles) the maximum 
current that we were able to put through the coil as a function of applied field together with 
the data of the 6-layer coil (in solid squares). 
For easy comparison with other published values, the right axis shows $J_c$ as a function 
of field (both coils were built with wires of the same diameters). 
The main result is that the critical current densities measured by transport are at least twice 
larger than those of the best samples without SiC (i.e: $J_c \sim 1 \times 10^4 A/cm^2$ 
at 7 T). Again due to heating problems at the contacts we could get reliable data only 
for applied magnetic fields above H=6.5 T. The solid and dotted lines 
are the $I_c$ obtained from magnetization measurements using the Bean model on 0.5 cm 
long pieces of the doped and un-doped wires, respectively.\\
   
In summary, we built a 6-layer, 4.5 cm-external diameter, 3.1 cm-long coil, by winding 
25 m of 1 mm-diameter powder-in-tube MgB$_2$ wire and subsequently hot isostatic 
pressing. At $T=4~$K and $H^{a}=1.25~$T it generates a field of 1.6 T (total field at 
the coil center 2.85 T). At $T=24.6~$K and $H^{a}=0$ the generated field was 
$H^{gen}(24.6~$K$)=1$ T. The performance of this compact coil satisfies the requirements 
for use in liquid-helium-free MRI systems.The developed method can be scaled to process coils
 with diameter about 1 meter.
We also explored the combination of HIPing with SiC doping and we found a significant 
improvement of $J_c$ for high fields.\\

\begin{figure}
\vspace{5cm}
\includegraphics[width=145mm]{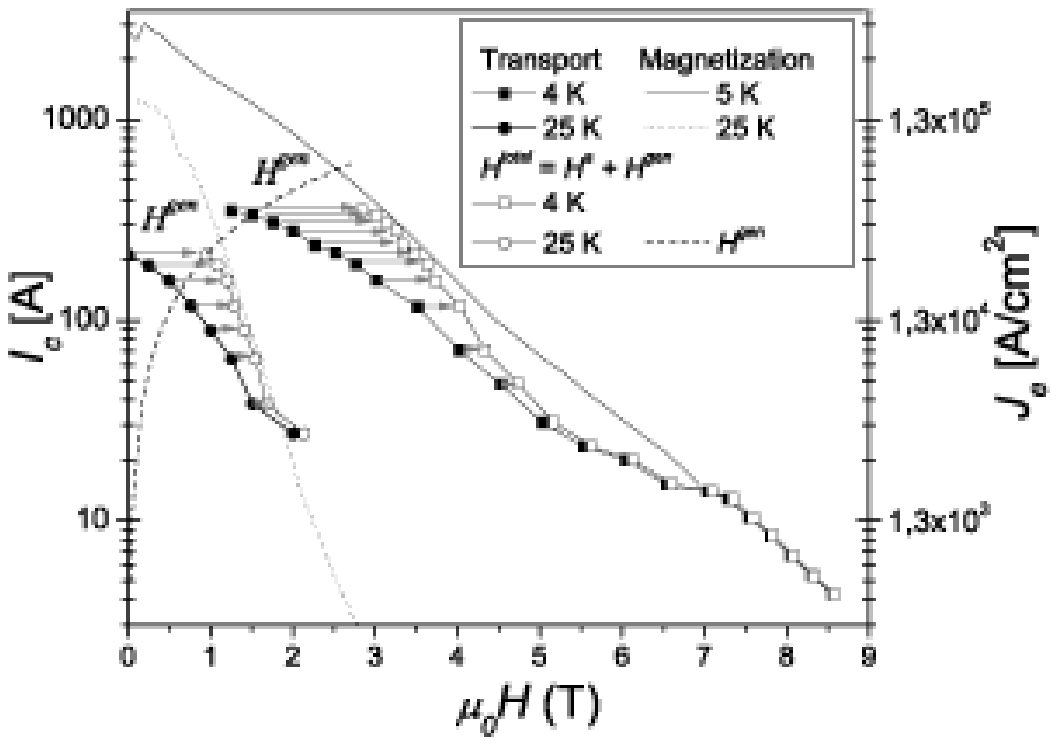}
\caption[]{Critical current $I_c$ (left axis) and engineering critical current density $J_e$
 (right axis) vs. magnetic field at $T=4~$K and 25 K of the 6-layer coil. The full symbols 
indicate the external field applied by the 9T magnet. The open symbols indicate the total 
field at the coil center. The difference (indicated by arrows) is the field $H^{gen}$ 
generated by the coil. Solid and dotted lines: magnetization data for a short piece of the coil wire. Dashed line: $I_c$ vs. $H^{gen}$ relation.}
\label{figure2}
\end{figure}

\begin{figure}
\vspace{5cm}
\includegraphics[width=145mm]{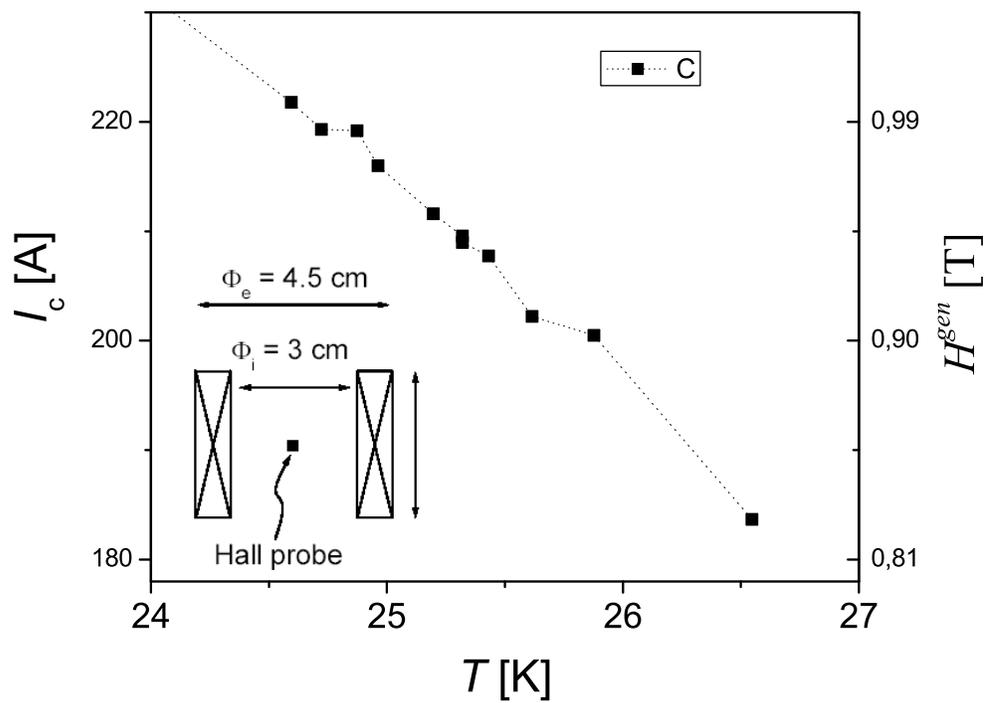}
\vspace{-0.5cm}
\caption[]{Critical current (left axis) and field generated by the coil (right axis) vs. 
temperature of the 6-layer coil.}
\vspace{0.2cm}
\label{figure3} 
\end{figure}

\begin{figure}
\vspace{5cm}
\includegraphics[width=145mm]{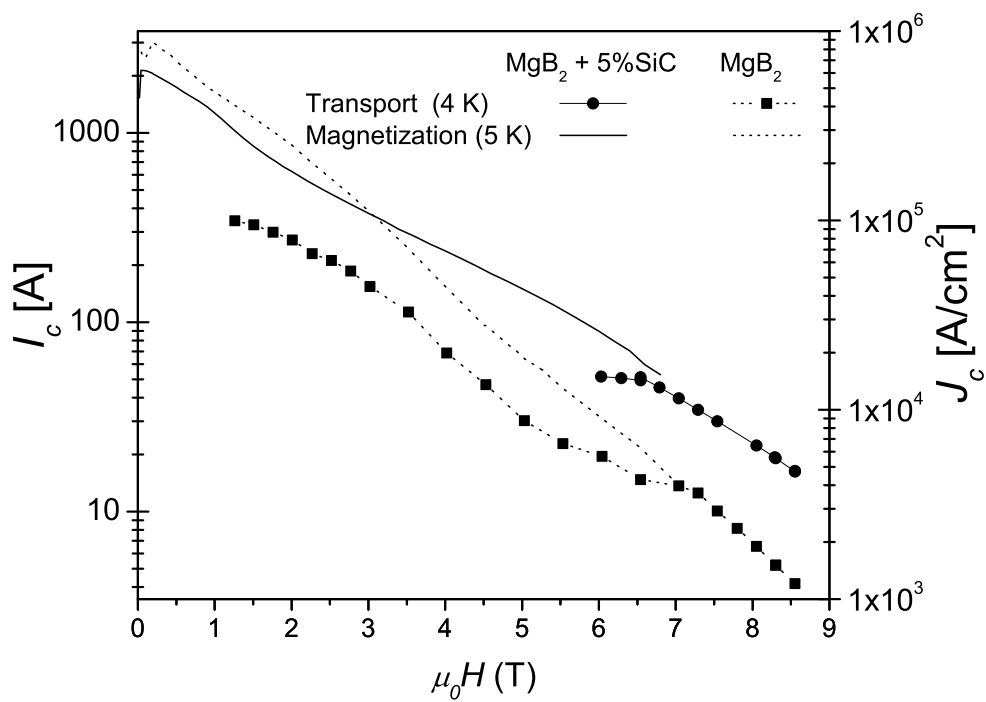}
\caption[]{Critical current $I_c$ (left axis) and critical current density $J_c$ (right axis) vs. magnetic field 
at $T=4~$K for the 1-layer SiC doped MgB$_2$ coil. The inset shows a photograph of the 
1-layer coil. For comparison we also included the 6-layer coil data. Solid and dotted lines:
 magnetization data 
for short pieces of the doped and un-doped wires, respectively.}
\label{figure4}
\end{figure}


\end{document}